\documentclass[prl,aps,twocolumn,floatfix,superscriptaddress,showpacs]{revtex4}

\usepackage{graphicx}
\usepackage{dcolumn}
\usepackage{amssymb}
\usepackage{amsmath}
\usepackage{latexsym}
\usepackage{verbatim}
\usepackage{color} 

\begin{document}

\title {
  Changes in the gradient percolation transition caused by an Allee effect
}

\author{Michael T.~Gastner}
\affiliation{Institute for the Chemistry and Biology of the Marine 
  Environment, Carl von Ossietzky Universit\"at, Carl-von-Ossietzky-Stra{\ss}e 
  9-11, 26111 Oldenburg, Germany}
\affiliation{Department of Mathematics, Complexity and Networks Progamme,
  Imperial College London, South Kensington Campus, London SW7 2AZ, United
  Kingdom}
\author{Beata Oborny}
\affiliation{Department of Plant Taxonomy and Ecology, Lor\'and E\"otv\"os
  University, P\'azm\'any P\'eter stny. 1/C, H-1117, Budapest, Hungary}
\author{Alexey B.~Ryabov}
\affiliation{Institute for the Chemistry and Biology of the Marine 
  Environment, Carl von Ossietzky Universit\"at, Carl-von-Ossietzky-Stra{\ss}e 
  9-11, 26111 Oldenburg, Germany}
\author{Bernd Blasius}
\affiliation{Institute for the Chemistry and Biology of the Marine 
  Environment, Carl von Ossietzky Universit\"at, Carl-von-Ossietzky-Stra{\ss}e
  9-11, 26111 Oldenburg, Germany}

\begin{abstract} 
  The establishment and spreading of biological populations depends 
  crucially on population growth at low densities.
  The Allee effect is a problem in those populations where the per-capita
  growth rate at low densities is reduced.
  We examine stochastic spatial models in which the reproduction rate 
  changes across a gradient $g$ so that the population undergoes a 
  2D-percolation transition.
  Without the Allee effect, the transition is continuous and the width $w$ of
  the hull scales as in conventional (i.e., uncorrelated) gradient 
  percolation, $w\propto g^{-0.57}$.
  However, with a strong Allee effect the transition is first order and 
  $w\propto g^{-0.26}$.
\end{abstract}
\maketitle	

It is not just human relationships which obey the rule ``two's company, three's 
a crowd.''
Negative density dependence, defined as a decrease of the per-capita
growth rate with increasing population density, is common among almost all 
species at high densities, where overcrowding and the depletion of resources
limit further growth.
The most common model for negative density dependence is the logistic
equation which assumes that the per-capita growth rate
decreases linearly with the population size $P$,
\begin{equation}
  \frac1P \frac{dP}{dt} = r\left(1-\frac PK\right),
  \label{logistic}
\end{equation}
where $t$ is time, $r$ is the intrinsic rate of increase, and $K$ the 
carrying capacity.

If $r,K>0$, Eq.~\ref{logistic} is characterized by a negative density 
dependence for all population sizes $P$.
For some small populations, however, a positive density dependence
can be observed.
The latter is called a demographic Allee effect, named after Warder Clyde 
Allee, who described it first and supported the theory with examples from 
various animal species from insects to mammals~\cite{Allee31}.
Small populations can suffer from reduced growth rates for various reasons.
Frequently, a collective behavior (e.g., defense against predators) becomes
inefficient when the group is small.
Additionally, small populations are less efficient in modifying the 
environment to their own benefit.
For example, plant individuals in aggregations can reduce frost or 
desiccation, but only when the density in the clump is sufficiently 
high~\cite{Brooker_etal08}.
To generalize density dependence, Volterra proposed to replace the right-hand 
side of Eq.~\ref{logistic} with a quadratic function of $P$~\cite{Volterra38},
\begin{equation}
  \frac1P \frac{dP}{dt} = -A+BP-CP^2,\;\;\;A,B,C>0.
  \label{Volterra}
\end{equation}
If $B^2-4AC > 0$, Eq.~\ref{Volterra} has two stable fixed points, unlike 
Eq.~\ref{logistic}, which has only one, so that the long-term behavior of 
Eq.~\ref{Volterra} depends on the initial population density.
If $P>(B-\sqrt{B^2-4AC})/2C$ at $t=0$, the population will approach a positive 
limit, whereas a smaller initial population will become extinct.
Several other formulations of the Allee effect have been suggested in the 
past decades, some including stochastic and spatial 
effects~\cite{HoltRoy,Johnson_etal06} (see chapter 3.5 in 
Ref.~\onlinecite{Courchamp_etal08} for a review).
They all have in common that a strongly positive density dependence 
accelerates the extinction of small populations.

\begin{figure}
  \begin{center}
    \includegraphics[width=8.6cm]{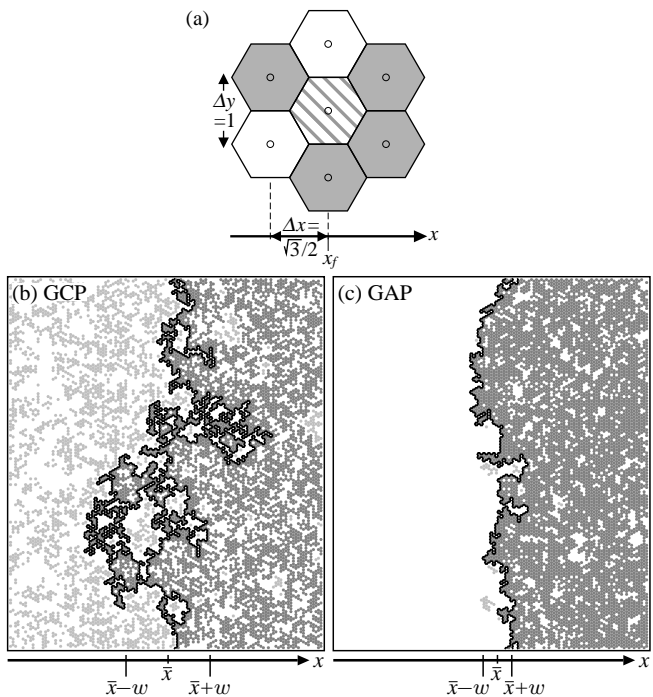}
    \caption{
      (a) Sites in the spatial models are placed in the centers of the 
          hexagons in a honeycomb lattice. 
          Gray cells represent populated ($A$), white cells vacant 
          ($\emptyset$) sites. 
          In this example, the focal site in the center has $k=4$ populated 
          neighbors.
          If this site is populated, it will die during a small time interval 
          $dt$ with probability equal to $dt$. 
          If, on the other hand, this site is vacant, it will become populated 
          with probability $(k/6)\,b(x_f)dt$ in the GCP or 
          $\left[\frac12 k(k-1)/15\right]b(x_f)dt$ in the GAP, where $b(x_f)$ 
          is the local birth attempt rate.
      (b) Typical snapshot of the GCP,
      (c) of the GAP.
          Dark gray: the largest populated cluster.
          Light gray: all other populated sites. 
          Black curve: percolation hull.
          The mean hull position $\bar{x}$ and the width of the fluctuations 
          $w$ are indicated at the bottom.
    }
    \label{lattice}
  \end{center}
\end{figure}

The work described here is motivated by the question: what are the 
consequences of an Allee effect on populations that live at a margin of a 
geographic range?
Because such populations usually have low densities, one can expect that it
matters greatly for the success of establishment and spreading if an Allee
effect is present or not~\cite{Lewis_Keitt_Newman_Clerc,Johnson_etal06}.
In this Letter, we investigate the situation near a geographic margin with 
two models where the density changes across space from low to high values.
We show that a strong Allee effect makes the percolation transition at the
margin discontinuous, 
causing scaling behavior different from previously studied types of
gradient percolation~\cite{Sapoval_etal85,Hansen_Boissin_Roux_Loscar}.

Our models are stochastic cellular automata whose local rules correspond to 
discretized versions of Eq.~\ref{logistic} or \ref{Volterra}.
Both cellular automata operate on a two-dimensional honeycomb lattice where 
the sites are either populated ($A$) or vacant ($\emptyset$, 
Fig.~\ref{lattice}a).
They can change their state by local death and birth events.
In both models, deaths are Poisson processes:
\begin{itemize}
\vspace{-.2cm}
\item{$A\rightarrow\emptyset$: A populated site becomes vacant with rate $1$.}
\end{itemize}
\vspace{-.2cm}
In our first model, the rate, with which a vacant site becomes populated by
a local birth event, is exactly proportional to the number of neighbors:
\begin{itemize}
\vspace{-.2cm}
\item{$A\rightarrow 2A$: A vacant site at position $(x,y)$ with $k$ populated 
  adjacent sites becomes itself populated at the rate $b(x)\cdot k/6$.}
\end{itemize}
\vspace{-.2cm}
The second model implements a local Allee effect by requiring at least one 
pair of neighbors for successful births.
The rule $A\rightarrow 2A$ is replaced with:
\begin{itemize}
\vspace{-.2cm}
\item{$2A\rightarrow 3A$: A vacant site at $(x,y)$ with $k$ neighbors 
  [i.e., $\frac12 k(k-1)$ pairs of neighbors] becomes populated with rate 
  $b(x)\cdot\frac12 k(k-1)/15$.}
\end{itemize}
\vspace{-.2cm}
The denominators 6 and 15 are the maximum number of neighbors and the maximum 
number of neighbor pairs, respectively.
Sites are updated in a random order with the rates stated above following the 
algorithm of Ref.~\onlinecite{Gastner_etal09}.

The function $b(x)$ can be interpreted as the rate with which an 
individual in column $x$ attempts to produce offspring on an adjacent site.
A birth attempt succeeds only if that site is vacant.
In the case of $2A\rightarrow 3A$, success further depends on a second 
neighbor adjacent to the newly born individual.
If $b(x)$ is a constant, then the first model is equivalent to a contact
process~\cite{Harris74}, and our second model becomes a special case of 
``Schl\"ogl's second model''~\cite{Schloegl_Grassberger_Prakash_Durrett_Liu}.

Our work differs from previous studies by assuming a constant gradient $g>0$ 
in the birth attempt rate, $b(x)=gx$.
Long-range gradients are important in ecology because the environmental
conditions can change gradually over distances larger than the distance of
dispersal within one generation (e.g., along a hillside or across geographic
latitudes).
We call $A\rightarrow 2A$ a gradient contact process 
(GCP)~\cite{Gastner_etal09, Oborny_etal09} and $2A\rightarrow 3A$ a gradient 
Allee process (GAP).
As long as the initial population density is sufficiently high in regions with
high birth rates, the population density reaches a steady state independent
of the initial conditions (see supplement).

Because $g>0$, the steady-state density of populated sites grows in both the 
GCP and the GAP as $x$, and hence $b$, increases.
At small $x$, the populated sites form small isolated patches (light gray 
sites in Figs.~\ref{lattice}b and \ref{lattice}c) whereas at large $x$ most 
populated sites belong to one large cluster (dark gray).
The curve along which the largest cluster touches the largest contiguous 
vacant area (black curve) is the percolation hull~\cite{Voss84,Sapoval_etal85}.
If the populated sites provide habitat or food for another species, the hull 
marks the borderline between the connected and fragmented occurrence of this
resource.
An example is a treeline across an altitudinal or latitudinal gradient.
Births and deaths cause the position and shape of the percolation hull to
fluctuate.
The average position of the hull $\bar{x}$ and the characteristic width of the
fluctuations $w$ depend on $g$ and the model (GCP versus GAP).
We compute $\bar{x}$ and $w$ as the mean and the standard deviation of the 
distribution of $x$-coordinates along the hull during several independent 
runs.

\begin{figure}
  \begin{center}
    \includegraphics[width=8.6cm]{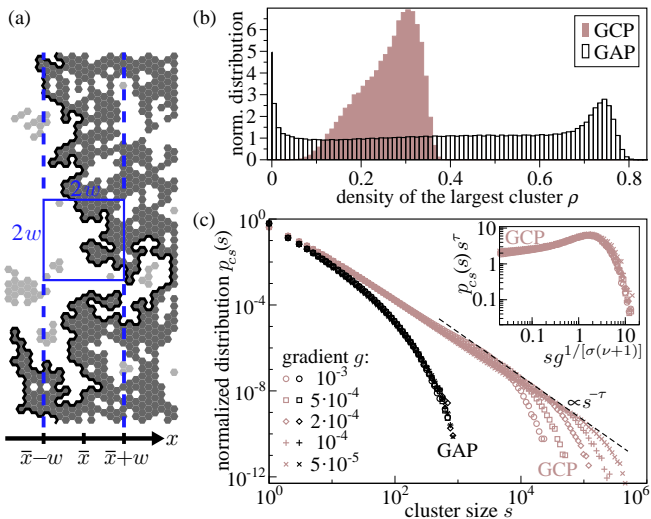}
    \caption{
      (a) To distinguish between a continuous and a first-order transition, 
      we investigate squares of dimension $2w\times 2w$ centered at $\bar{x}$.
      We measure the fraction of sites $\rho$ that belong to the cluster
      covering the largest area within the square. 
      (Typically, though not necessarily, this is the largest cluster on the 
      entire lattice, shown in dark gray).
      (b) From data of several independent runs, we obtain the probability
      distribution for $\rho$, represented as a histogram.
      We show data for $g=5\cdot 10^{-5}$.
      The smaller the gradient, the more weight is concentrated in the two 
      peaks of the GAP distribution.
      (c) The distribution of cluster sizes $s$ in the stripe $|x-\bar{x}|<w$.
      We include clusters which are at least partially in this stripe,
      but exclude the system's largest cluster.
      The dashed line $\propto s^{-\tau}$ is the tangent to the GCP distribution.
      The inset shows the data collapse for the GCP.
      The exponents are those of standard 2D-percolation:
      $\tau=\frac{187}{91}$, $\sigma=\frac{36}{91}$, 
      $\nu=\frac{4}{3}$~\cite{StaufferAharony91}.
    }
    \label{1_vs_2}
  \end{center}
\end{figure}

In Fig.~\ref{lattice}(b), the number of sites in the GCP's largest 
cluster increases gradually from left to right.
The increase is much more abrupt in the GAP (Fig.~\ref{lattice}c) which 
generates fewer isolated clusters.
This impression can be confirmed by looking at local densities in 
the transition region near $\bar{x}$.
Because $w$ is the relevant length scale in this region, we investigate
subsystems located between $\bar{x}-w$ and $\bar{x}+w$ (Fig.~\ref{1_vs_2}a).
We determine the cluster that has the largest number of sites $N$ within a
$2w\times 2w$ square.
In both models, the mean of $N$ scales approximately as 
$w^{D_f}$ with $D_f=91/48$,
the fractal dimension of the incipient infinite cluster in standard (i.e.,
uncorrelated, gradient-free, nondirected) two-dimensional 
percolation~\cite{StaufferAharony91} (see supplement).
But the distribution of the $N$ sites is different in the two models.

To see this quantitatively, let us define the cluster density $\rho$
as $N$ divided by the number of all (populated or vacant) sites in the square.
The distributions of $\rho$, aggregated over independent runs of the GCP
and the GAP, at different $y$-coordinates and at different times, are 
shown in Fig.~\ref{1_vs_2}b.
The GCP distribution has a single peak at intermediate densities whereas the 
GAP has two local maxima, one at zero and another one at high density.
In analogy to thermodynamics, where a bimodal probability distribution of an
order parameter is an indication of a first-order phase 
transition~\cite{LeeKosterlitz90}, percolation in the GAP can be interpreted 
as a first-order transition between two steady states of either zero or of a 
positive density.
Thus only a population larger than a critical density is able to grow and, 
when this density is exceeded, the cluster size grows abruptly, reminiscent 
of recent reports of ``explosive
percolation''~\cite{Achlioptas_Araujo_Moreira}, but generated here by a 
purely local rule.

In some explosive percolation models, cluster sizes were recently shown to 
follow power-law distributions, casting doubts on whether the transition is 
truly first order~\cite{Radicchi_Ziff_daCosta}.
In the GAP, by contrast, the cluster size distribution $p_{cs}(s)$ in the 
stripe $|x-\bar{x}|<w$ gives further evidence in support of a 
first-order transition (Fig.~\ref{1_vs_2}c).
While the GCP distribution follows the scaling behavior expected for 
two-dimensional percolation with a continuous transition 
$p_{cs}(s) = s^{-\tau}f_{cs}(s g^{1/[\sigma(\nu+1)]})$, the GAP 
distribution does not show indications of scaling.
There is neither a power-law decay nor a dependence on the gradient even for
large cluster sizes.
Instead, consistent with a first-order transition, there is a characteristic
size for the GAP clusters.

\begin{figure}
  \begin{center}
    \includegraphics[width=8.6cm]{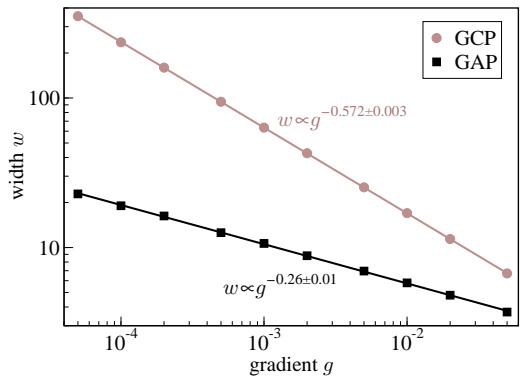}
    \caption{
      The hull width $w$ as a function of the gradient $g$.
      The lines are least-squares fits to the data.
      Error bars are smaller than the symbol sizes.
    }
    \label{width}
  \end{center}
\end{figure}

Another percolation property affected by the Allee effect is the scaling 
relation between $w$ and $g$ (Fig.~\ref{width}).
For the GCP, we find $w\propto g^{-a_\text{GCP}}$, $a_\text{GCP}=0.572(3)$ 
(95\% confidence interval).
The GAP width also follows a power law, but with a smaller exponent 
$a_\text{GAP}=0.26(1)$.
(The same exponents are observed if the hull is replaced with the accessible
external perimeter~\cite{Grossman87}, see supplement.)
The exponent $a_\text{GCP}$ is in agreement with $\nu/(\nu+1)=4/7$, which
was derived for uncorrelated percolation based on the divergence of the
correlation length~\cite{Sapoval_etal85}.
There is no analogous relation for $a_\text{GAP}$, because the correlation 
length is finite at a first-order transition.
The result that $w$, nevertheless, scales with $g$ in the GAP is surprising, 
considering that scaling in stochastic gradient models has so far always
been linked to divergent correlation 
lengths~\cite{Hansen_Boissin_Roux_Loscar}.

\begin{figure}
  \begin{center}
    \includegraphics[width=8.6cm]{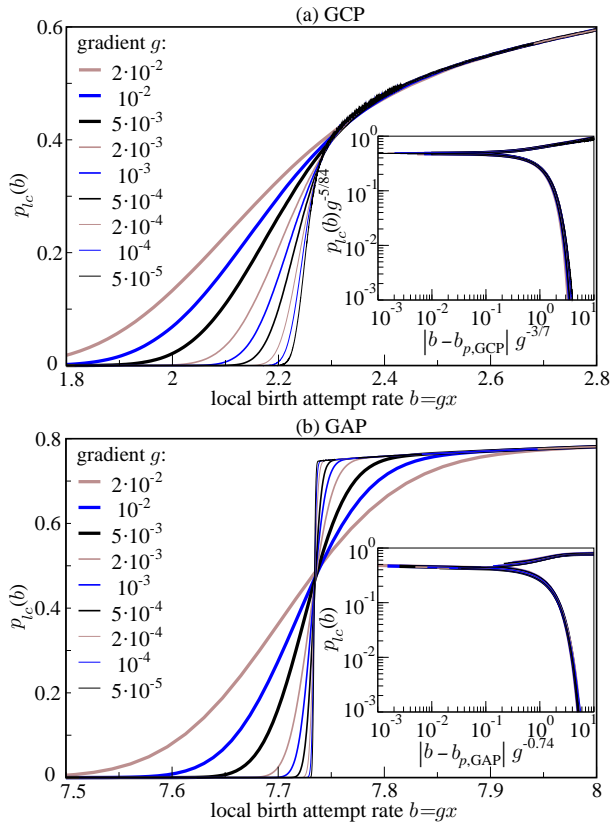}
    \caption{The probability $p_{lc}$ that a site belongs to the largest 
      cluster as a function of the site's birth attempt rate $b$ in (a) the 
      GCP, (b) the GAP for various gradients $g$.
      Insets: The functions collapse if the coordinates are rescaled.
      In both insets, the same nine data sets are shown as in the main panels.
    }
    \label{p_b}
  \end{center}
\end{figure}

Although the spatial width of the hull increases with decreasing $g$, 
the transition zone becomes, in terms of the birth rate $b$, more confined.
This becomes clear by plotting $p_{lc}(b)$, the
probability that a site with birth rate $b$ belongs to the largest 
cluster (Fig.~\ref{p_b}). 
In both models, $p_{lc}(b)$ approaches a limiting function as $g\to0^+$ with a
sharp increase at the percolation thresholds $b_{p,\text{GCP}}=2.260(1)$ and 
$b_{p,\text{GAP}}=7.7340(3)$.
For finite $g$, $p_{lc}(b)$ obeys the finite-size scaling
$p_{lc}(b,g) = g^{c} f_{lc}(|b-b_p|g^{a-1})$, where $a$ is the hull width 
exponent ($a_\text{GCP}$ or $a_\text{GAP}$) and $f_{lc}$ is a model-dependent 
scaling function.
In the GCP, we expect $c=\beta/(\nu+1) = 5/84$.
For the GAP, however, the first-order transition demands that $p_{lc}$ has a
discontinuity in the limit $g\to 0^+$ and therefore $c_\text{GAP}=0$.
The insets of Fig.~\ref{p_b} show 
a remarkable 
data collapse for the anticipated exponents.

Why is percolation in the GAP unconventional?
Let us denote by $n(b,t)$ the probability that a site with birth attempt rate 
$b$ is populated at time $t$.
The mean-field equations for $n$ to lowest order in $g$ are
\begin{equation}
  \text{GCP: }\;  
  \frac{\partial n}{\partial t} = -n + b(1-n)n +
  \frac{g^2}4 b(1-n)\frac{\partial^2 n}{\partial b^2},
    \label{GCP}
\end{equation}
\begin{align}
  \text{GAP: }\;  
  \frac{\partial n}{\partial t} = &-n + b(1-n)n^2 + \nonumber\\
  &\frac{g^2}{10} b (1-n)
  \left[5 n \frac{\partial^2 n}{\partial b^2} - 
        \left(\frac{\partial n}{\partial b}\right)^2 \right],\label{GAP}
\end{align}
where $b(x)=gx$ (see supplement for the derivation and numerical solutions).
If $g=0$, Eq.~\ref{GCP} turns into the logistic equation~(\ref{logistic}) 
with $r=b-1$, $P=bKn/r$, and Eq.~\ref{GAP} becomes Eq.~\ref{Volterra} with 
$P=n$, $A=1$, $B=C=b$.
For $g\to 0^+$, Eq.~\ref{GCP} has a continuous stationary solution: 
$n=\max(0,1-1/b)$.
In Eq.~\ref{GAP}, on the other hand, $n$ develops a discontinuity in the limit
$g\to 0^+$ where it suddenly jumps from zero to $n\approx 0.77$.
If the percolation threshold is at a probability $n_p = 0.5$, as in
uncorrelated honeycomb lattices~\cite{StaufferAharony91}, 
the GCP hull is at a position where $n(b)$ varies smoothly.
The GAP hull, by contrast, lies at a position where $n(b)$ changes abruptly.

In summary, the GCP and the GAP behave fundamentally differently near the 
margin of the populated range.
The GCP, a model without any Allee effect, possesses the same characteristic 
features as previously reported for uncorrelated gradient 
percolation~\cite{Gastner_etal09}.
The Allee effect in the GAP changes the situation drastically: the 
percolation transition is first order and the hull width diverges more 
slowly for $g\to 0^+$.

We thank H.~J. Jensen, G.~Pruessner, O.~Peters, R.~Dickman, A.~O. Parry, 
A.~Windus, G.~Cs\'anyi and S.~Mughal for helpful discussions. 
M.~T.~G. gratefully acknowledges support by the German Volkswagen Foundation 
and Imperial College.
B.~O. is supported by the Hungarian National Science Foundation (Grant 
No.~OTKA K61534) and the Santa Fe Institute International Program.
Calculations were performed on the GOLEM I cluster of the Universit\"at
Oldenburg.

\end{document}